\documentclass[aps,prd,reprint,preprintnumbers,nofootinbib,amsmath,amssymb,floatfix]{revtex4-2}

\usepackage{graphicx}
\usepackage{bm}
\usepackage{mathtools}
\usepackage[colorlinks=true,linkcolor=blue,citecolor=blue,urlcolor=blue]{hyperref}
\hypersetup{
 pdftitle={Resonant electromagnetic leptogenesis with helicity-resolved transport},
 pdfauthor={Rin Takada}
}

\DeclareMathOperator{\diag}{diag}
\newcommand{\dd}{\mathrm d}
\newcommand{\mEM}{\tilde m^{\rm EM}}
\newcommand{\kmatch}{\kappa_{\rm match}}
\newcommand{\kN}{\kappa_N}
\newcommand{\Tr}{\operatorname{Tr}}
\newcommand{\1}{\bm 1}
\newcommand{\MtildeBenchmark}{3.97\times10^{-2}}
\newcommand{\VacuumWidthBenchmark}{1.04\times10^{-10}}
\newcommand{\DecayParameterBenchmark}{74.2}
\newcommand{\DecayOnlyPeakSplitting}{1.29\times10^{-11}}
\newcommand{\DecayOnlyPeakYB}{5.50\times10^{-5}}
\newcommand{\DecayOnlyObservedSplitting}{1.68\times10^{-5}}
\newcommand{\DecayOnlyObservedRelativeSplitting}{1.68\times10^{-8}}
\newcommand{\DecayPlusScatteringObservedSplitting}{2.34\times10^{-5}}
\newcommand{\DecayPlusScatteringObservedRelativeSplitting}{2.34\times10^{-8}}
\newcommand{\ObservedCrossingShiftPercent}{39.4\%}
\newcommand{\SphaleronSwitchTemperature}{142.74}
\newcommand{\SphaleronFreezeTemperature}{131.55}
\newcommand{\GammaOverHAtMOne}{1.28\times10^{4}}
\newcommand{\GammaOverHAtTc}{4.75\times10^{2}}
\newcommand{\MaxCoherencePlus}{1.59\times10^{-2}}
\newcommand{\MaxCoherenceMinus}{1.80\times10^{-2}}
\newcommand{\DecayPlusScatteringPeakSplitting}{1.83\times10^{-11}}
\newcommand{\DecayPlusScatteringPeakYB}{8.44\times10^{-5}}
\newcommand{\PoleChoicePeakDifferencePercent}{53.5\%}

\begin{document}

\preprint{RESCEU-7/26}

\title{Resonant electromagnetic leptogenesis with coherent heavy-neutrino evolution}

\author{Rin Takada}
\email{takada-rin@resceu.s.u-tokyo.ac.jp}
\affiliation{Research Center for the Early Universe (RESCEU), Graduate School of Science,
The University of Tokyo, 7-3-1 Hongo, Bunkyo, Tokyo 113-0033, Japan}
\date{July 17, 2026}

\begin{abstract}
We study TeV-scale electromagnetic leptogenesis generated by the gauge-invariant neutrino-dipole
operators $\mathcal O_{NB}$ and $\mathcal O_{NW}$. The benchmark Wilson coefficients are specified
at the renormalization scale $\kN=M_1=1\,\mathrm{TeV}$ and related to their values at
$\kmatch=3\,\mathrm{TeV}$ through the coupled one-loop $\nu$SMEFT evolution. The collision network
contains the symmetric-phase $1\leftrightarrow3$ and crossed $2\leftrightarrow2$ processes generated
by the linear field-strength vertices and the VEV-induced two-body $\gamma$, $Z$, and $W$ channels
through the electroweak crossover. We use a factorized leading-moment, momentum-averaged Markovian treatment
for two $2\times2$ heavy-flavor density matrices, one for each helicity. Pole-resummed two-state
effective couplings enter the lepton and antilepton collision tensors, while the off-diagonal
density matrices describe coherent evolution and collision damping. Three resolved flavor charges
are coupled to a finite-rate sphaleron equation, yielding the frozen baryon yield $Y_B^{\rm FO}$.
We work in an effective-coupling, leading-moment approximation; intermediate-shell contributions
and thermal dispersive self-energies in the deep-resonant regime are not evaluated.
Maxwell-Boltzmann statistics, massless bath particles in the symmetric-phase phase space, and a
mean-field crossover profile are used.

For the benchmark $\mEM_1=\MtildeBenchmark\,\mathrm{eV}$, the decay-only pole
prescription, in which only decay contributions enter the absorptive pole matrix, gives the
locally optimized positive near-resonant maximum $Y_B^{\rm FO}=+\DecayOnlyPeakYB$ at
$\Delta M=\DecayOnlyPeakSplitting\,\mathrm{GeV}$. Adding the thermally averaged
crossed-scattering moment to the pole matrix defines the decay-plus-scattering prescription and
shifts the maximum to $Y_B^{\rm FO}=+\DecayPlusScatteringPeakYB$ at
$\Delta M=\DecayPlusScatteringPeakSplitting\,\mathrm{GeV}$. The corresponding signed
mass-splitting curves cross the observed value $Y_B^{\rm obs}$ at
$\Delta M=\DecayOnlyObservedSplitting\,\mathrm{GeV}$ and
$\DecayPlusScatteringObservedSplitting\,\mathrm{GeV}$.
\end{abstract}

\maketitle

\section{EFT framework and benchmark}
\label{sec:eft}

A lepton asymmetry generated by heavy Majorana fermions can be converted into baryon number by
electroweak sphalerons~\cite{Sakharov1967,KlinkhamerManton1984,FukugitaYanagida1986}.
Electromagnetic leptogenesis replaces the ordinary neutrino-Yukawa decay vertex by transition-dipole
interactions~\cite{BellKayserLaw2008}; a TeV-scale realization with resonant enhancement was proposed
in Ref.~\cite{Choudhury2012}. Above the electroweak crossover we use the $\nu$SMEFT
dipoles~\cite{BellKayserLaw2008,ChalaTitov2020}
\begin{align}
\mathcal O_{NB,\alpha i}
 &\coloneqq(\bar L_\alpha\sigma^{\mu\nu}P_RN_i)\tilde H B_{\mu\nu},
\label{eq:ONB}\\
\mathcal O_{NW,\alpha i}
 &\coloneqq(\bar L_\alpha\sigma^{\mu\nu}\tau^AP_RN_i)\tilde H W^A_{\mu\nu},
\label{eq:ONW}
\end{align}
where $\tau^A$ are the Pauli matrices and
$\sigma^{\mu\nu}\coloneqq\mathrm{i}[\gamma^\mu,\gamma^\nu]/2$. The effective Lagrangian is
\begin{equation}
-\mathcal L_{\nu{\rm SMEFT}}
 \supset C_{NB,\alpha i}\mathcal O_{NB,\alpha i}
    + C_{NW,\alpha i}\mathcal O_{NW,\alpha i}+\mathrm{h.c.},
\label{eq:EFTlag}
\end{equation}
where $C_{NX}$ has mass dimension $-2$. We consider two Majorana states,
\begin{equation}
M_2\coloneqq M_1+\Delta M,\qquad M_1=1\,\mathrm{TeV}.
\label{eq:masses}
\end{equation}

Motivated by the softly broken $Z_2$ model of Ref.~\cite{Choudhury2012}, we adopt the
following factorized flavor structure at the renormalization scale $\kN=M_1$, with the phases in
Eq.~\eqref{eq:texture} fixed to the quoted benchmark values:
\begin{equation}
C_{NX,\alpha i}
\coloneqq c_X y_{\Sigma,\alpha}y_{H,i}^{\ast},
\qquad
X=B,W,
\label{eq:factorized}
\end{equation}
with
\begin{align}
y_H
&\coloneqq(1.4,\,1.4\mathrm{e}^{-0.7\mathrm{i}})^{\intercal},
\notag\\
y_\Sigma
&\coloneqq(1.4,\,1.12\mathrm{e}^{-0.3\mathrm{i}},
\,0.84\mathrm{e}^{-1.1\mathrm{i}})^{\intercal},
\label{eq:texture}
\end{align}
and
\begin{align}
|C_{NB,e1}(\kN)|
&=1.041\times10^{-12}\,\mathrm{GeV}^{-2},
\notag\\
|C_{NW,e1}(\kN)|
&=1.809\times10^{-12}\,\mathrm{GeV}^{-2}.
\label{eq:Cinput}
\end{align}
The vector $y_H$ in Eq.~\eqref{eq:texture} is a
flavor-texture parameter and is not identified with the renormalizable $Y_\nu$. The ratio
\begin{equation}
\frac{C_{NW}(\kN)}{C_{NB}(\kN)}=+1.739
\label{eq:Cratioinput}
\end{equation}
is a real-positive EFT input at $\kN$.
The calculation uses one unrounded value of $C_{NB}$ and one unrounded real-positive ratio,
from which $C_{NW}$ is derived; the three displayed benchmark numbers are rounded separately.

In the Pauli-matrix convention of Eqs.~\eqref{eq:ONB} and \eqref{eq:ONW}, we retain the
gauge terms and the top-Yukawa trace in the one-loop equations
\cite{Datta2021,ChalaTitov2020,ArduMarcano2024},
\begin{align}
16\pi^2\kappa\frac{\dd C_{NB}}{\dd\kappa}
={}&\left(\frac{91}{12}g'^2-\frac94g^2+3y_t^2\right)C_{NB}\notag\\
&-\frac92gg'C_{NW},
\label{eq:RGEB}\\
16\pi^2\kappa\frac{\dd C_{NW}}{\dd\kappa}
={}&\left(-\frac34g'^2-\frac{11}{12}g^2+3y_t^2\right)C_{NW}\notag\\
&-\frac32gg'C_{NB}.
\label{eq:RGEW}
\end{align}
We neglect charged-lepton, down-quark, and neutrino-Yukawa terms in
Eqs.~\eqref{eq:RGEB} and \eqref{eq:RGEW}. The coupled evolution maps the benchmark inputs of
Eqs.~\eqref{eq:Cinput} and \eqref{eq:Cratioinput} to
\begin{align}
|C_{NB,e1}(\kmatch)|&=1.043\times10^{-12}\,\mathrm{GeV}^{-2},\notag\\
|C_{NW,e1}(\kmatch)|&=1.827\times10^{-12}\,\mathrm{GeV}^{-2}.
\label{eq:RGEresult}
\end{align}
The label $\kmatch$ denotes a renormalization reference scale; it is not used as a
kinematic cutoff on the reaction densities.
For the numerical evolution, the one-loop SM beta functions are initialized at
$\kappa_{\rm input}=150\,\mathrm{GeV}$ with
$(g',g,g_s,y_t)=(0.357,0.652,1.17,0.93)$.

After electroweak symmetry breaking,
\begin{align}
\mu^B_{\alpha i}(T)&\coloneqq\frac{v(T)}{\sqrt2}C_{NB,\alpha i},\notag\\
\mu^3_{\alpha i}(T)&\coloneqq\frac{v(T)}{\sqrt2}C_{NW,\alpha i},\notag\\
\mu^W_{\alpha i}(T)&\coloneqq v(T)C_{NW,\alpha i},\notag\\
\mu^\gamma_{\alpha i}(T)&\coloneqq c_W\mu^B_{\alpha i}(T)
+s_W\mu^3_{\alpha i}(T),\notag\\
\mu^Z_{\alpha i}(T)&\coloneqq-s_W\mu^B_{\alpha i}(T)
+c_W\mu^3_{\alpha i}(T).
\label{eq:brokenmu}
\end{align}
The vacuum-width parameter is
\begin{align}
\mEM_i&\coloneqq v^2M_i\sum_{\alpha,V}R_V(0)|\mu^V_{\alpha i}(0)|^2,\notag\\
\Gamma_i^{\rm vac}&\coloneqq\frac{M_i^2}{2\pi v^2}\mEM_i,
\label{eq:mtilde}
\end{align}
where
$R_V\coloneqq(1-r_V)^2(1+r_V/2)$ and $r_V\coloneqq m_V^2/M_i^2$. Here
$v=246\,\mathrm{GeV}$ and
$\mu^V(0)$ denote physical zero-temperature quantities; they are distinct from the mean-field
profile $v(T)$ used below and specified in Eq.~\eqref{eq:vT}. This is the dipole-width convention; the conventional Yukawa relation contains $8\pi$ instead of
$2\pi$. The benchmark has
$\mEM_1=\MtildeBenchmark\,\mathrm{eV}$,
$\Gamma_1^{\rm vac}=\VacuumWidthBenchmark\,\mathrm{GeV}$, and
$\Gamma_1^{\rm vac}/H(M_1)=\DecayParameterBenchmark$.

\section{Collision tensors}
\label{sec:collision}

\subsection{Symmetric- and broken-phase channels}

We use $z\coloneqq M_1/T$. In the symmetric phase, the Higgs doublet remains a dynamical bath field.
The linear field-strength terms generate
\begin{align}
N_i&\leftrightarrow L_\alpha+\tilde H+X,\notag\\
N_i+X&\leftrightarrow L_\alpha+\tilde H,\qquad
N_i+\tilde H^\dagger\leftrightarrow L_\alpha+X,\notag\\
N_i+\bar L_\alpha&\leftrightarrow \tilde H+X,\qquad X=B,W,
\label{eq:symprocesses}
\end{align}
and their CP conjugates. In the shorthand channel labels below, $H$ denotes the crossed Higgs
state associated with the $\tilde H$ field. With massless bath particles, one weak component and
one charge direction give
\begin{equation}
\Gamma_X=\frac{M^5}{512\pi^3}|C_{NX}|^2,
\qquad X=B,W.
\end{equation}
The two weak components and the CP-conjugate channel therefore give
\begin{equation}
\Gamma_{1\leftrightarrow3,X}^{\rm CP\ sum}
=\frac{M^5}{128\pi^3}|C_{NX}|^2.
\label{eq:3body}
\end{equation}
The corresponding crossed-channel cross sections, with $|C_{NX}|^2$ removed, are averaged over
the initial $N$ spin; the $NX$ channel is additionally averaged over the two transverse gauge
polarizations, while the chiral $\bar L$ state is not spin averaged:
\begin{align}
\sigma_{NX}(s)&=\frac{s-M^2}{32\pi},\notag\\
\sigma_{NH}(s)&=\frac{s(s+M^2)}{8\pi(s-M^2)},\notag\\
\sigma_{N\bar L}(s)&=\frac{s+2M^2}{24\pi}.
\label{eq:crosssections}
\end{align}
Their Maxwell-Boltzmann reaction densities are~\cite{Buchmuller2002}
\begin{align}
\gamma_a(T)&\coloneqq\frac{T}{64\pi^4}\int_{M^2}^{\infty}\dd s\,\sqrt{s}\,
\hat\sigma_a(s)K_1(\sqrt{s}/T),\notag\\
\hat\sigma_a(s)&\coloneqq\frac{2(s-M^2)^2}{s}\sigma_a(s),
\label{eq:reactiondensity}
\end{align}
and the coupling-stripped reduced scattering rate for one charge direction is
\begin{align}
r_{2\leftrightarrow2}(T)&\coloneqq
\frac{4\gamma_{NX}+2\gamma_{NH}+2\gamma_{N\bar L}}{n_N^{\rm eq}},\notag\\
n_N^{\rm eq}(T)&\coloneqq\frac{M^2TK_2(M/T)}{2\pi^2}.
\label{eq:rsym}
\end{align}
Here $n_N^{\rm eq}$ is the Maxwell-Boltzmann density for one heavy-neutrino spin state.
The factors $4,2,2$ restore the two initial gauge polarizations in
the $NX$ channel and count the two weak components in all three channels; the $N\bar L$ channel
contains only the interacting chiral helicity. The corresponding reduced decay rate is
$r_{1\leftrightarrow3}\coloneqq M^5[K_1(z)/K_2(z)]/(256\pi^3)$.

Below $T_c=159.5\,\mathrm{GeV}$ we use
\begin{align}
v(T)^2&\coloneqq\frac{c_T}{\lambda_H}(T_c^2-T^2),\notag\\
c_T&\coloneqq\frac{3g^2+g'^2}{16}+\frac{y_t^2}{4}
+\frac{\lambda_H}{2}=0.3687,
\label{eq:vT}
\end{align}
where $\lambda_H\coloneqq m_h^2/(2v^2)$ and $m_h=125.25\,\mathrm{GeV}$.
We set $v(T)=0$ above $T_c$. This mean-field interpolation, anchored at the lattice crossover
temperature~\cite{DOnofrio2014,DOnofrio2016}, is used only over the transport interval below
$T_c$ and is not extrapolated to $T=0$. The VEV-induced coupling-stripped reduced rate is
\begin{equation}
r_V(T)\coloneqq\frac{M^3}{4\pi}R_V(T)\frac{K_1(z)}{K_2(z)},
\label{eq:rV}
\end{equation}
which enters Eq.~\eqref{eq:dimensionlessh} together with the temperature-dependent couplings in
Eq.~\eqref{eq:brokenmu}.

\subsection{Pole resummation and helicity-resolved collision tensors}

For each collision channel $a$, define
\begin{equation}
h^a_{\alpha i}(T)\coloneqq\sqrt{\frac{16\pi r_a^{\rm coll}(T)}{M_1}}\,
g^a_{\alpha i}(T),
\label{eq:dimensionlessh}
\end{equation}
where $g^a$ is $C_{NB}$, $\sqrt{3}\,C_{NW}$, or a broken-phase dipole matrix. The quantities
$r_a^{\rm coll}$ are coupling-stripped reduced rates: they have mass dimension $5$ for the
symmetric-phase channels, where $g^a$ has mass dimension $-2$, and mass dimension $3$ for the
broken-phase channels, where $g^a$ has mass dimension $-1$. Thus $h^a$ is dimensionless.
These reduced rates contain thermal time dilation and the crossed $2\leftrightarrow2$ network.
In Eq.~\eqref{eq:dimensionlessh},
the $B$ and $W$ matrices entering $g^a$ are $C_{NB}$ and $\sqrt{3}\,C_{NW}$, respectively.
The common phase-space kinematics is evaluated at $M=M_1$; terms of relative order
$\Delta M/M_1$ in the collision moments are neglected, while the state masses are retained in
the pole factors and the dispersive Hamiltonian.

The decay contribution to the vacuum-form pole matrix is
\begin{equation}
A^{\rm D}_{ij}(T)\coloneqq\frac{1}{16\pi}
\sum_{d\in\mathrm{dec},\alpha}\hat h^d_{\alpha i}(T)
\hat h^{d*}_{\alpha j}(T),
\label{eq:Aij}
\end{equation}
where $\hat h^d$ uses the rest-frame $1\to3$ and $\gamma,Z,W$ decay rates. We evaluate two
choices: $A^{\rm pole}\coloneqq A^{\rm D}$ for decay only, and
$A^{\rm pole}\coloneqq A^{\rm D}+A^{\rm S}$ for decay plus scattering, where $A^{\rm S}$ is built from
the thermally averaged crossed-scattering moment. In both choices, decay and scattering remain in
the collision tensors.

For two states, the effective amplitudes in the pole-mass basis are
\begin{equation}
\bm h^a_{\alpha i}\coloneqq h^a_{\alpha i}
-\mathrm{i}h^a_{\alpha j}
\frac{M_i\left(M_iA^{\rm pole}_{ij}+M_jA^{\rm pole}_{ji}\right)}
{M_i^2-M_j^2+2\mathrm{i}M_i^2A^{\rm pole}_{jj}},\qquad j\ne i,
\label{eq:resummedh}
\end{equation}
with the CP-conjugate amplitudes $\bm h^{c}$ obtained by replacing the bare amplitudes
$h\to h^{\ast}$ before resummation~\cite{Pilaftsis1997,PilaftsisUnderwood2004,DevCovariant2014}.

In the factorized leading-moment approximation, we assign the Maxwell-Boltzmann-averaged
helicity weights
\begin{align}
w_\lambda(z)&\coloneqq\frac{1+\lambda\bar v(z)}2,\qquad \lambda=\pm1,\notag\\
\bar v(z)&\coloneqq\frac{2(z+1)\mathrm{e}^{-z}}{z^2K_2(z)}.
\label{eq:helicityweight}
\end{align}
The scalar $r_a^{\rm coll}$ in Eq.~\eqref{eq:dimensionlessh} is initial-spin averaged. The
assigned fixed-helicity rates therefore contain $w_\lambda/(1/2)=2w_\lambda$. The lepton and
antilepton tensors
are
\begin{align}
[\Gamma^{\ell}_{\lambda\alpha}]_{ij}
 &\coloneqq2w_\lambda\frac{\sqrt{M_iM_j}}{16\pi}
  \sum_a\bm h^{a*}_{\alpha i}\bm h^a_{\alpha j},
\label{eq:Glep}\\
[\Gamma^{\bar\ell}_{\lambda\alpha}]_{ij}
 &\coloneqq2w_{-\lambda}\frac{\sqrt{M_iM_j}}{16\pi}
  \sum_a\bm h^{ca*}_{\alpha i}\bm h^{ca}_{\alpha j}.
\label{eq:Ganti}
\end{align}
The helicity decomposition and density-matrix organization follow the leading-moment treatments
of Refs.~\cite{DevCovariant2014,DevKB2015,GhiglieriLaine2017}. The diagonal heavy-flavor
entries contain the pole-mixing correction, whereas contractions with off-diagonal densities
contain coherent contributions. Intermediate-shell terms beyond this leading-moment prescription
are not evaluated in the deep-resonant regime~\cite{DevKB2015,Kartavtsev2016}. After the tensors
are constructed in the pole basis, the moment equations are covariant under simultaneous
similarity transformations of the heavy-flavor matrices.

\section{Helicity-resolved transport and sphaleron freeze-out}
\label{sec:transport}

Let $\rho_\lambda$ be the $2\times2$ heavy-flavor yield matrix for helicity $\lambda$. With
$s\coloneqq2\pi^2g_*T^3/45$, where $g_*$ is the effective number of relativistic degrees of freedom, the
equilibrium yield of one Majorana flavor is
$Y_N^{\rm eq}\coloneqq2n_N^{\rm eq}/s$, and the yield per helicity is
$Y_h^{\rm eq}\coloneqq n_N^{\rm eq}/s=Y_N^{\rm eq}/2$.
We use a momentum-averaged Markovian leading-gradient ansatz based on the flavor-covariant and
helicity-resolved kinetic structures of Refs.~\cite{DevCovariant2014,DevKB2015,GhiglieriLaine2017}.
The equations are
\begin{align}
\frac{\dd\rho_\lambda}{\dd z}
={}&-\frac{\mathrm{i}}{H(T)z}[\Omega_N,\rho_\lambda]
-\frac{1}{2H(T)z}\{\Gamma_\lambda,\rho_\lambda-Y_h^{\rm eq}\1\}
\notag\\
&+\frac{Y_h^{\rm eq}}{H(T)z}
\sum_\alpha\Delta\Gamma_{\lambda\alpha}\,\xi_\alpha,
\label{eq:rhoQKE}
\end{align}
where
\begin{equation}
\Gamma_\lambda\coloneqq\sum_\alpha(\Gamma^\ell_{\lambda\alpha}+\Gamma^{\bar\ell}_{\lambda\alpha}),
\qquad
\Delta\Gamma_{\lambda\alpha}\coloneqq\Gamma^\ell_{\lambda\alpha}-\Gamma^{\bar\ell}_{\lambda\alpha},
\label{eq:ratecomb}
\end{equation}
and the lepton-doublet chemical potentials $\xi_\alpha$ are specified in
Eq.~\eqref{eq:chemicalpotential} below. The unpolarized damping diagnostic quoted below is
$\Gamma_N^{\rm av}\coloneqq\frac12\sum_\lambda\Gamma_\lambda$. The normalized coherence shown in
Fig.~\ref{fig:evolution} is
$\mathsf C_\lambda\coloneqq|\rho_{\lambda,12}|/
\sqrt{\rho_{\lambda,11}\rho_{\lambda,22}}$. The Hamiltonian is
\begin{align}
\Omega_N&\coloneqq\diag(-\Delta\omega/2,+\Delta\omega/2),\notag\\
\Delta\omega&\coloneqq\frac{M_2^2-M_1^2}{2M_1}\frac{K_1(z)}{K_2(z)}.
\label{eq:Hamiltonian}
\end{align}
The anticommutator damps both populations and heavy-flavor coherence. All source and
damping terms are constructed from the rate tensors in Eqs.~\eqref{eq:Glep} and
\eqref{eq:Ganti}; no additional CP-odd source or decoherence rate is included. Equation~\eqref{eq:Hamiltonian}
retains the vacuum mass splitting and omits the thermal dispersive self-energy, including its
off-diagonal heavy-flavor component.

The flavor charges
\begin{equation}
Y_{\Delta_\alpha}\coloneqq\frac13Y_B-Y_{L_\alpha}
\label{eq:YDelta}
\end{equation}
obey
\begin{align}
\frac{\dd Y_{\Delta_\alpha}}{\dd z}
={}&-\frac{1}{H(T)z}\sum_\lambda
\Tr\bigl[\Delta\Gamma_{\lambda\alpha}
(\rho_\lambda-Y_h^{\rm eq}\1)\bigr]
\notag\\
&+\frac{Y_h^{\rm eq}}{H(T)z}\sum_\lambda
\Tr\bigl(\Gamma^\ell_{\lambda\alpha}+\Gamma^{\bar\ell}_{\lambda\alpha}\bigr)\xi_\alpha.
\label{eq:YDeltaQKE}
\end{align}
The second line is the linearized inverse-reaction washout. Ref.~\cite{Eijima2017} uses the
opposite charge convention, $Y_{L_\alpha}-Y_B/3=-Y_{\Delta_\alpha}$. After this conversion,
the lepton-doublet chemical potentials are
\begin{equation}
\xi_\alpha\coloneqq\frac{\mu_\alpha}{T}
=\frac{s}{T^3}\left[-\bar\omega_{\alpha\beta}(T)Y_{\Delta_\beta}
+\bar\omega_B(T)Y_B\right],
\label{eq:chemicalpotential}
\end{equation}
where $\bar\omega\coloneqq T^2\omega$ has entries~\cite{Eijima2017}
\begin{align}
\bar\omega_{\alpha\beta}
&\coloneqq a\,\delta_{\alpha\beta}+b\,(1-\delta_{\alpha\beta}),\notag\\
a&\coloneqq\frac{22(15x^2+44)}{9(17x^2+44)},\qquad
b\coloneqq\frac{8(3x^2+22)}{9(17x^2+44)},
\label{eq:ab}
\end{align}
and
$\bar\omega_B\coloneqq4(27x^2+77)/[9(17x^2+44)]$, with
$x\coloneqq v(T)/T$.

For $Y_B\coloneqq n_B/s$, the finite-rate baryon equation is
\begin{equation}
\frac{\dd Y_B}{\dd z}
=-\frac{\Gamma_B(T)}{H(T)z}
\left[Y_B-\chi(T)\sum_\alpha Y_{\Delta_\alpha}\right],
\label{eq:sphaleronQKE}
\end{equation}
where
\begin{align}
\chi(T)&\coloneqq4\frac{27x^2+77}{333x^2+869},
\label{eq:chi}\\
\Gamma_B(T)&\coloneqq9\frac{869+333x^2}{792+306x^2}
\frac{\Gamma_{\rm diff}(T)}{T^3},
\label{eq:GammaB}\\
\frac{\Gamma_{\rm diff}}{T^4}
&=\begin{cases}
18\alpha_W^5,&T\geqslant T_c,\\
\exp[-147.7+0.83T/\mathrm{GeV}],&T<T_c.
\end{cases}
\label{eq:Gammadiff}
\end{align}
Here $\alpha_W\coloneqq g^2/(4\pi)$ with $g=0.652$. The relaxation equation and susceptibilities follow
Ref.~\cite{Eijima2017}; the diffusion-rate fit and crossover temperatures use
Refs.~\cite{DOnofrio2014,DOnofrio2016}. For $\Gamma_B/H>10^4$, the fast-sphaleron relation
$Y_B=\chi\sum_\alpha Y_{\Delta_\alpha}$ is used. The finite-rate equation is evolved below
$T=\SphaleronSwitchTemperature\,\mathrm{GeV}$; $\Gamma_B/H=1$ occurs at
$T=\SphaleronFreezeTemperature\,\mathrm{GeV}$. The system is evolved from $T=M_1$ to
$100\,\mathrm{GeV}$. For $\Delta M\geqslant 3\times10^{-10}\,\mathrm{GeV}$, the rapidly
oscillating off-diagonal
moments are treated in the adiabatic limit; the direct and adiabatic results differ by less than
$7\times10^{-6}$ in their overlap region.

\section{Results}
\label{sec:numerics}

We take $g_{\ast}=106.75$. The unpolarized rate satisfies
$(\Gamma_N^{\rm av})_{11}/H=\GammaOverHAtMOne$ at $T=M_1$ and $\GammaOverHAtTc$ at $T=T_c$.

\begin{figure}[t]
\includegraphics[width=\columnwidth]{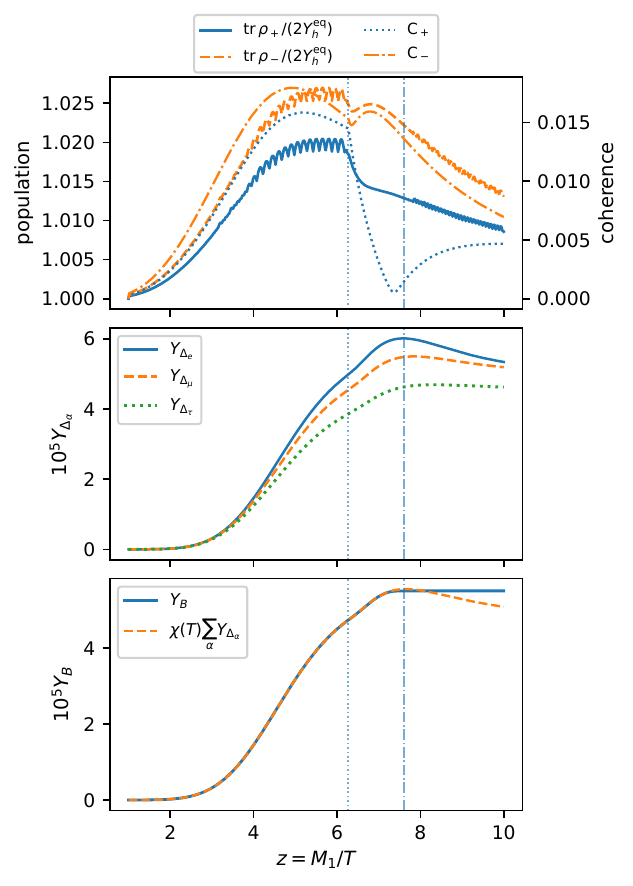}
\caption{Heavy-neutrino helicity populations and coherences (top), resolved flavor charges (middle),
and the finite-rate baryon solution (bottom) at the decay-only near-resonant splitting in
Eq.~\eqref{eq:nearresult}, with $\mEM_1=\MtildeBenchmark\,\mathrm{eV}$ and a thermal initial
heavy-neutrino density. Here
$\mathsf C_\lambda\coloneqq|\rho_{\lambda,12}|/
\sqrt{\rho_{\lambda,11}\rho_{\lambda,22}}$,
$Y_{\Delta_\alpha}\coloneqq Y_B/3-Y_{L_\alpha}$, and $Y_B\coloneqq n_B/s$.
The vertical dotted and dash-dotted lines mark $T_c$ and the temperature
$\Gamma_B/H=1$, $T=\SphaleronFreezeTemperature\,\mathrm{GeV}$, respectively.}
\label{fig:evolution}
\end{figure}

Fig.~\ref{fig:evolution} shows the two helicity populations, their normalized heavy-flavor
coherences, the resolved charges, and the baryon yield at the EFT benchmark
$\mEM_1=\MtildeBenchmark\,\mathrm{eV}$. For decay only, the locally optimized positive
near-resonant maximum is
\begin{equation}
\Delta M=\DecayOnlyPeakSplitting\,\mathrm{GeV},\qquad
Y_B^{\rm FO}=+\DecayOnlyPeakYB.
\label{eq:nearresult}
\end{equation}
The decay-plus-scattering pole prescription gives
$\Delta M=\DecayPlusScatteringPeakSplitting\,\mathrm{GeV}$ and
$Y_B^{\rm FO}=+\DecayPlusScatteringPeakYB$. The two maxima differ by $\PoleChoicePeakDifferencePercent$. The maximal normalized coherences are $\MaxCoherencePlus$ and
$\MaxCoherenceMinus$ for the two helicities. The dashed curve in the lower panel is the
instantaneous equilibrium value $\chi\sum_\alpha Y_{\Delta_\alpha}$; the finite-rate solution
departs from it during sphaleron freeze-out and approaches a constant below the freeze-out interval.

\begin{figure}[t]
\includegraphics[width=\columnwidth]{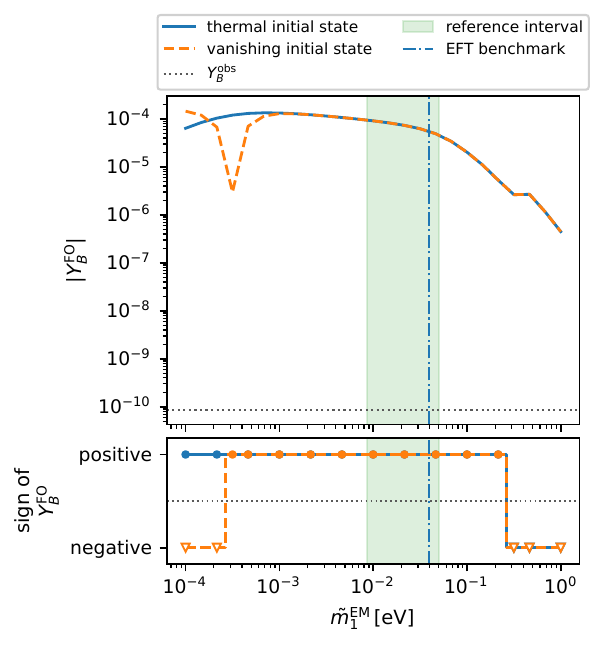}
\caption{Frozen baryon yield as a function of $\mEM_1$ for decay only at the fixed splitting
$\Delta M=\DecayOnlyPeakSplitting\,\mathrm{GeV}$. The upper panel shows $|Y_B^{\rm FO}|$ for
thermal and vanishing initial heavy-neutrino densities; the lower panel labels the sign as positive
or negative. Filled circles and open downward triangles denote positive and negative points,
respectively. The green band is the neutrino-mass reference interval in
Eq.~\eqref{eq:greenband}; the dash-dotted vertical line is the EFT benchmark.}
\label{fig:mtilde}
\end{figure}

Fig.~\ref{fig:mtilde} scans the overall dipole strength using decay only at the fixed splitting
$\Delta M=\DecayOnlyPeakSplitting\,\mathrm{GeV}$. The upper panel shows the magnitude
$|Y_B^{\rm FO}|$, while the lower panel shows its sign. The green interval
\begin{equation}
8.6\times10^{-3}\,\mathrm{eV}\lesssim\mEM_1\lesssim5.0\times10^{-2}\,\mathrm{eV}
\label{eq:greenband}
\end{equation}
is a visual reference interval constructed from the square roots of the measured neutrino
mass-squared splittings~\cite{NuFIT2024}. It is not a direct experimental bound on $\mEM$:
$\mEM$ parameterizes the dipole width, while ultraviolet completions can connect the same dipole
operator to light-neutrino masses~\cite{BellKayserLaw2008}.
The strong symmetric-phase reactions erase the dependence on the initial heavy-neutrino abundance
in the neutrino-mass reference interval of $\mEM_1$. Both initial conditions give a positive frozen
baryon asymmetry throughout that interval.

\begin{figure}[t]
\includegraphics[width=\columnwidth]{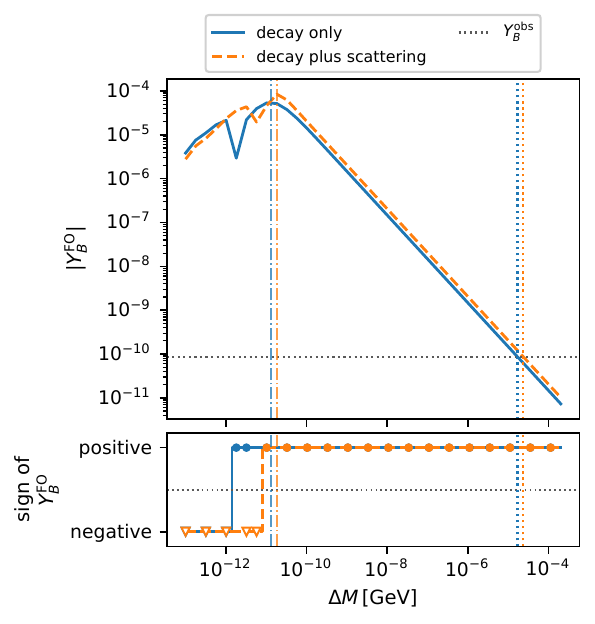}
\caption{Frozen baryon yield versus the heavy-neutrino mass splitting. The solid blue and dashed
orange curves correspond to decay only and decay plus scattering in the pole matrix, respectively.
The upper panel shows $|Y_B^{\rm FO}|$ and the lower panel shows the sign of $Y_B^{\rm FO}$.
The horizontal dotted line marks $Y_B^{\rm obs}$. The
dash-dotted vertical lines mark the near-resonant maxima, and the dotted vertical lines mark the
positive crossings in Eq.~\eqref{eq:observedresult}.}
\label{fig:dM}
\end{figure}

The mass-splitting curves are shown in Fig.~\ref{fig:dM}.
The sign reversals in Figs.~\ref{fig:mtilde} and \ref{fig:dM} occur without changing the benchmark
CP phases and reflect changes in the net source after washout over the thermal history.
Converting the positive
baryon-to-photon ratio inferred from Planck to the yield $Y_B\coloneqq n_B/s$ gives
$Y_B^{\rm obs}=+8.7\times10^{-11}$~\cite{Planck2018}. The positive crossings are
\begin{align}
\Delta M_{\rm obs}^{\mathrm{D}}&=\DecayOnlyObservedSplitting\,\mathrm{GeV},&
\frac{\Delta M_{\rm obs}^{\mathrm{D}}}{M_1}&=\DecayOnlyObservedRelativeSplitting,\notag\\
\Delta M_{\rm obs}^{\mathrm{D+S}}&=\DecayPlusScatteringObservedSplitting\,\mathrm{GeV},&
\frac{\Delta M_{\rm obs}^{\mathrm{D+S}}}{M_1}&=\DecayPlusScatteringObservedRelativeSplitting.
\label{eq:observedresult}
\end{align}
Here ``D'' denotes decay only, while ``D+S'' denotes decay plus scattering in the pole matrix.
Both crossings satisfy $Y_B^{\rm FO}=+8.7\times10^{-11}$ and differ by
$\ObservedCrossingShiftPercent$.

\section{Conclusions}
\label{sec:conclusion}

The leading-moment system retains both heavy flavors, both helicities,
non-diagonal rate tensors, coherent Hamiltonian evolution, collision damping, resolved lepton
charges, and the finite-rate sphaleron response. Intermediate-shell contributions beyond this
prescription are not evaluated in the deep-resonant regime~\cite{DevKB2015,Kartavtsev2016}.

The neutrino-mass reference interval gives positive $Y_B^{\rm FO}$, and both pole-matrix choices
yield a positive solution matching the observed asymmetry.
The decay-only and decay-plus-scattering prescriptions differ by $\PoleChoicePeakDifferencePercent$
at the positive near-resonant maximum and by $\ObservedCrossingShiftPercent$ at the positive
observed-yield crossing.
These variations show the sensitivity of the leading-moment result to whether the thermally
averaged scattering moment is included in the pole matrix.

We use Maxwell-Boltzmann statistics, massless bath particles in
Eqs.~\eqref{eq:3body}-\eqref{eq:rsym}, a mean-field interpolation for $v(T)$, and the vacuum
dispersive Hamiltonian in Eq.~\eqref{eq:Hamiltonian}. The omitted off-diagonal thermal dispersive
term enters at the same coupling order as the absorptive matrix and may shift the deep-resonant
peak by an order-one amount. The quoted peak positions and heights therefore apply to the
vacuum-dispersive prescription; the observed-yield crossings lie far outside this regime.

We keep $g_*=106.75$ fixed, neglecting its temperature dependence and the heavy-neutrino
contribution, which changes $H$ and $s/T^3$ at the few-percent level. The piecewise diffusion-rate
fit is discontinuous at $T_c$, but the join lies in the fast-sphaleron regime and does not affect the
frozen yield. The mean-field $v(T)$ profile is a crossover approximation, not a zero-temperature
extrapolation. The Wilson coefficients and the small renormalized mass splitting are EFT inputs.

\begin{acknowledgments}
The author is grateful to K. Hotokezaka, R. Jinno, and R. Namba for helpful comments.
\end{acknowledgments}

\appendix

\section{Symmetric-phase phase space}
\label{app:symmetric}

For fixed lepton-flavor and heavy-neutrino indices, the relevant
interactions are
\begin{align}
-\mathcal L \supset {}&
C_{NB}
(\bar L_a\sigma^{\mu\nu}P_RN)\tilde H_a B_{\mu\nu}
\nonumber\\
&+
C_{NW}
[\bar L_a\sigma^{\mu\nu}(\tau^A)_{ab}P_RN]
\tilde H_b W^A_{\mu\nu}
+\mathrm{h.c.}
\end{align}
For a fixed weak and gauge component, the group-theory factor may be stripped off. We denote
the incoming heavy-neutrino momentum by $P$, and the outgoing lepton, Higgs, and gauge-boson
momenta by $p$, $k$, and $q$, respectively, with $P=p+k+q$. Suppressing the fixed flavor indices,
the common Lorentz amplitude for $X=B,W$ is
\begin{equation}
\mathcal M_X
\coloneqq2C_{NX}\bar u(p)\sigma^{\mu\nu}q_\mu P_Ru(P)
\varepsilon_\nu^\ast(q).
\end{equation}
With massless final states,
\begin{equation}
\frac12\sum_{\rm spin,pol}|\mathcal M_X|^2
=16|C_{NX}|^2(P\cdot q)(p\cdot q).
\end{equation}
Writing $2p\cdot q=xM^2$ and $2k\cdot q=yM^2$, with
$x,y\geqslant0$ and $x+y\leqslant1$, gives
$\Gamma_X=|C_{NX}|^2M^5/(512\pi^3)$. The two weak components and the
CP-conjugate channel give Eq.~\eqref{eq:3body}. Crossing yields
\begin{align}
|\mathcal M_X(NX\to LH)|^2&=-2|C_{NX}|^2(s-M^2)t,\notag\\
|\mathcal M_X(NH\to LX)|^2&=4|C_{NX}|^2s(s+t),\notag\\
|\mathcal M_X(N\bar L\to HX)|^2&=-4|C_{NX}|^2t(s+t).
\end{align}
For $NX\to LH$, averaging over the two initial transverse gauge polarizations supplies an
additional factor $1/2$. The resulting cross sections give Eq.~\eqref{eq:crosssections} after
integration over $-(s-M^2)\leqslant t\leqslant0$.

\section{Helicity projection and pole-basis covariance}
\label{app:helicity}

For a right-chiral interaction, the on-shell helicity projector gives
$(1+\lambda\|\bm p\|/E)/2$. By Maxwell-Boltzmann averaging,
\begin{equation}
\left\langle\frac{\|\bm p\|}{E}\right\rangle
\coloneqq\frac{\int_0^\infty\dd p\,p^3E^{-1}\mathrm{e}^{-E/T}}
{\int_0^\infty\dd p\,p^2\mathrm{e}^{-E/T}}
=\frac{2(z+1)\mathrm{e}^{-z}}{z^2K_2(z)}.
\end{equation}
In the factorized leading-moment prescription, the scalar rates in
Eq.~\eqref{eq:dimensionlessh} are combined with this averaged projector. Since those rates
are averaged over the two initial spin states, the fixed-helicity factors are
$2w_\lambda=1+\lambda\bar v$ and $2w_{-\lambda}=1-\lambda\bar v$ in
Eqs.~\eqref{eq:Glep} and \eqref{eq:Ganti}. The antilepton channel carries the opposite helicity
weight because the chiral projector acts on the CP-conjugate amplitude
\cite{GhiglieriLaine2017}.

The resummed amplitudes are evaluated in the pole-mass basis and converted to rate tensors before
any basis test. Under a simultaneous similarity transformation of the moment variables,
\begin{equation}
\rho_\lambda\to U^\dagger\rho_\lambda U,
\qquad
\Omega_N\to U^\dagger\Omega_NU,
\qquad
\Gamma_{\lambda\alpha}\to U^\dagger\Gamma_{\lambda\alpha}U,
\end{equation}
every term in Eqs.~\eqref{eq:rhoQKE} and \eqref{eq:YDeltaQKE} transforms identically and
the source traces are invariant. This covariance applies to the pole-basis moment tensors with
the corresponding flavor metric~\cite{DevCovariant2014}.

\end{document}